\documentclass[10 pt, letter]{article}
\usepackage{graphicx}
\usepackage{caption}
\usepackage{amsmath}
\usepackage{amsthm}
\usepackage{enumerate}
\usepackage{subfig}
\usepackage{float}
\usepackage{amsfonts}
\usepackage{amssymb}
\usepackage{anysize}
\usepackage{helvet}
\usepackage{setspace}
\usepackage{multirow}
\usepackage{tabularx}
\usepackage{tabulary}
\usepackage{longtable}
\usepackage{float}
\usepackage{mathtools}
 \def\ds{\displaystyle}
\title{\bf Force-Depending Radiation Reaction study in an undulator device}
\author{Gustavo V. L\'opez\footnote{gulopez@cencar.udg.mx}~ and Jorge Lizarraga~\\ \\
 Departamento de F\'{i}sica, Universidad de Guadalajara,\\
 Blvd. Marcelino Garc\'{i}a Barragan y Calzada Ol\'{i}mpica, \\�44200 Guadalajara, Jalisco, Mexico}
\begin{document}
\maketitle

\begin{abstract}
\noindent
The effect of force-depending radiation reaction on charge motion traveling inside an undulator is studied using the 
new force approach for radiation reaction. The effect on the dynamics of a  charged particle is determined with the 
hope that this one can be measured experimentally and can be determined whether or not this approach points on the right
direction to understand the nature of radiation reaction. 
\end{abstract}
\newpage
\section{ Introduction}
Radiation reaction due to radiation emission of electromagnetic waves of an accelerated particle [1,2] has been a hard topic for 
more than two centuries, and Abraham-Lorentz-Dirac approaches [3,4,5] to this phenomenon resulted really unsatisfactory  when
these are applied to observed experimental phenomenon in the nature since, for zero external forces on the charged particle, still
there exist some, so called, pre-acceleration of the charged particle (but this would imply radiation !, according to Maxwell's equations).
This observation was pointed out in [6], and the proposition that radiation reaction force must be a function of the external force was made there.  In this paper, we apply this approach to the dynamical motion of a relativistic single charged particle traveling in an undulator, and we make  the comparison of the dynamical motion with and without radiation reaction force, with the idea of to see and to quantify a possible difference.
 \section{Equation of Motion} 
  Consider the motion of a single charged particle inside an undulator magnet [7] of length $L$ with a magnetic field given by
  \begin{equation}\label{Bfield}
  {\bf B}(y)=\bigl(0,0,B_0\sin\frac{2\pi y}{\lambda_u}\bigr),
  \end{equation}
where $\lambda_u$ is the wave length of the undulator, defined by the periodicity of the ferromagnetic elements of the undulator (magnetic elements with 
alternating N-S pole arranged), and having its symmetry along the y-axis. The external force on the charged particle is (CGS units) [1]
\begin{equation}\label{ef}
{\bf F}=\frac{q}{c}{\bf v}\times{\bf B},
\end{equation}
where $q$ and ${\bf v}$ are the charge and the velocity of the particle, and $c$ is the speed of light. The radiation reaction force term is given by [8]
\begin{equation}
{\bf F}_{rad}=-\frac{\ds q^2|\aleph {\bf F}|^2{\bf v}}{\ds 4\pi m^2c^3\gamma^2v^2}\int_{\Omega}\frac{|{\bf e}_1\sin\theta_1-{\bf e}_2\beta\sin\theta_2|^2}
{1-\beta\cos\theta)^5}~d\Omega,
\end{equation} 
where $m$, $v$, $\beta$ are the mass, the speed, and the normalized speed ($\beta=v/c$) of the charged particle, $\gamma$ is the usual relativistic factor
($\gamma=(1-\beta^2)^{-1/2}$), and $\aleph$ is a matrix defined as
\begin{equation}
\aleph=\begin{pmatrix}1-\beta_x^2&-\beta_x\beta_y&-\beta_x\beta_z\\ \\
-\beta_y\beta_x& 1-\beta_y^2&-\beta_y\beta_z\\�\\
-\beta_z\beta_x&-\beta_z\beta_y&1-\beta_z^2\end{pmatrix},
\end{equation}
with $\beta_i$ for $i=x,y,z$ being the normalized components of the velocity of the charged particle. Since the problem is similar to a circular motion, the radiation 
reaction force is just
\begin{equation}
{\bf F}_{rad}=-\frac{\lambda_0F^2}{v^2\gamma^2}~{\bf v},
\end{equation} 
where $\lambda_0$ is defined as
\begin{equation}
\lambda_0=\frac{\ds 2q^2}{\ds 3m^2c^3}.
\end{equation}
Therefore, the relativistic equation of motion is 
\begin{equation}
\frac{d(\gamma{\bf v})}{dt}={\bf F}-\frac{\lambda_0F^2}{v^2\gamma^2}~{\bf v}. 
\end{equation}
In terms of the variable $\vec\beta={\bf v}/c$ and after making the differentiation of $\gamma$, the equation for $\vec\beta$ is
\begin{equation}\label{ebeta}
\dot{\vec\beta}=\frac{1}{mc\gamma}\aleph{\bf F}-\frac{\lambda_0F^2}{mc^2\beta^2\gamma^3}~{\aleph\vec\beta}\ .
\end{equation}
Substituting (\ref{Bfield}) and (\ref{ef}) in the above expression and after some rearrangements, one gets the following dynamical system
\begin{subequations}
\begin{eqnarray}
\dot x&=&\beta_xc\\ \nonumber \\
\dot \beta_x&=&A\beta_y\cos\frac{2\pi y}{\lambda_u}-B\biggl[(1-\beta_x^2)\beta_x-\beta_x\beta_y^2-\beta_x\beta_z^2\biggr]
\end{eqnarray}
\end{subequations}
\begin{subequations}
\begin{eqnarray}
\dot y&=&\beta_yc\\�\nonumber\\
\dot\beta_y&=&-A\beta_x\cos\frac{2\pi y}{\lambda_u}-B\biggl[-\beta_y\beta_x^2+(1-\beta_y^2)\beta_y-\beta_y\beta_z^2\biggr]
\end{eqnarray}
\end{subequations}
\begin{subequations}
\begin{eqnarray}
\dot z&=&\beta_zc\\�\nonumber\\
\dot\beta_z&=&-B\biggl[-\beta_z\beta_x^2-\beta_z\beta_y^2+(1-\beta_z^2)\beta_z\biggr],
\end{eqnarray} 
\end{subequations}
where the constants $A$ and $B$ are defined as
\begin{equation}
A=\frac{\ds cqB_0}{\ds mc^2\gamma}, \quad\quad\hbox{and}\quad\quad B=\frac{\ds \lambda_0F^2}{\ds mc^2\beta^2\gamma^3},
\end{equation}
and the magnitude of the force is 
\begin{equation}
F=qB_0\sqrt{\beta_x^2+\beta_y^2}~\cos\frac{2\pi y}{\lambda_u }.
\end{equation}
This dynamical systems is defined in the space $\Omega=\Re^3\times [-1,1]^3$, and the critical points in this space is the set points $\bigl\{({\bf x},\vec\beta)\in\Omega~|~\vec\beta={\vec 0}\bigr\}$. The linear matrix around each critical point is
\begin{equation}
D_{\bf x}=\begin{pmatrix}
0&0&0&c&0&0\\ \\
0&0&0&0&c&0\\�\\
0&0&0&0&0&c\\�\\
0&-a&0&-b&0&0\\ \\
0&-a&0& 0&-b&0\\ \\
0&-a&0& 0&0&-b\end{pmatrix},
\end{equation}
where $a$ and $b$ are given by
\begin{subequations} 
\begin{equation}
a=-\frac{2A\pi}{\lambda_u}\sin\frac{2\pi y}{\lambda_u}-\frac{2\lambda_0q^2B_0^2}{mc^2}\cos(2\pi y/\lambda_u)\sin(2\pi y/\lambda_u)
\end{equation}
and
\begin{equation}
b=\lambda_0q^2B_0^2\cos^2(2\pi y/\lambda_u)/mc^2.
\end{equation} 
\end{subequations}
The set of eigenvalues of this matrix, $\{\lambda_i\}_{i=1,\dots,6}$ are such that $Re(\lambda_i)\le 0$ for $i=1,\dots,6$  (in fact, one has that $Re(\lambda_i)\sim -b$
or $Re(\lambda_i)\sim -b\pm\sqrt{b^2-4ac}~$). Therefore, our dynamical system is stable for any ${\bf x}\in \Re^3$, and this stability is due to radiation reaction force, 
as one could have expected.
\section{Results}
Using a typical parameters for a free electron laser (FEL) 
\begin{equation}
\lambda_u=4~cm, \quad B_0=2000~Gauss,\quad q_e=-4.803\times10^{-10}esu,\quad m_e=9.109\times 10^{-34}gr,\quad c\approx 3\times 10^{10}cm/s,
\end{equation}
and the initial conditions of the electron as
\begin{equation}
x(0)=y(0)=z(0)=0,\quad \beta_x(0)=\beta_z(0)=0.00001, \quad \beta_y(0)=0.99,
\end{equation}
the figure below shows the difference (without and with radiation reaction force) of the normalized components of the velocity of the electron as a function of the length of the undulator (L). The continuous curve is the analytical fix to the behavior of this difference, which is of the form 
\begin{equation}
\Delta\beta_i=a_iL^2e^{b_iL}, \quad i=x,y,
\end{equation}
where $a_x= 1.94268\times 10^{-16}cm^{-2}$, $a_y=2.50485\times 10^{-17}cm^{-2}$, $b_x=1.40958\times 10^{-05}cm^{-1}$, and $b_y=2.26834\times 10^{-06}cm^{-1}$.  
\begin{figure}[H]
\includegraphics[scale=0.50,angle=0]{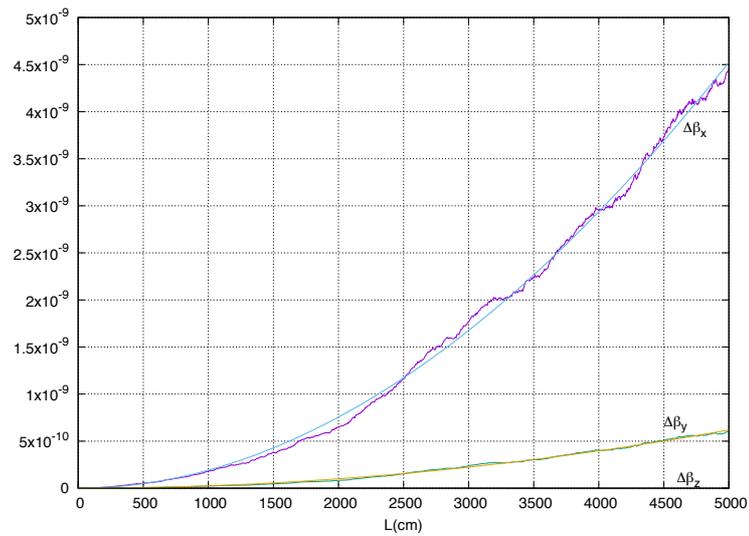}
\centering
    \caption{ Difference of the normalized  velocity components as a function of the undulator length. }
\end{figure}
\newpage
\section{Conclusions}
The force-depending radiation reaction approach was used to study the dynamical motion of a single changed particle in an undulator device.  We have shown
that there is a small difference in the dynamics of the particle when radiation reaction force is taken into account, and  this difference increases exponentially as a function of the undulator length $L$. Although, the difference on velocity components is quite small even for long undulators, the results suggest that this difference could be measured experimentally, and in turns, one can see whether or not this
approach points in the right direction to understand radiation reaction force. 

\newpage\noindent
{\Large\bf References}\\�\\
1. J.D. Jackson, {\it Classical Electrodynamics}, Chapter 17, John Wiley\&Sons, Inc., (1962).\\�\\
2. L.D. Landau and F.M. Lifshitz, {\it The classical Theory of Fields}, Chapter 9, Butterworth Heinemann, (2002).\\�\\
3. M. Abraham and R. Recker, {\it Electricity and Magnetism}, Blackie, London, (1937).\\�\\
4. H.A. Lorentz, {\it The theory of electron}, Dover, (1952).\\�\\
5. P.A.M. Dirac, Proc. Roy. Soc. London A{\bf 167}, (1938), 148.\\�\\
6. G.V. L\'opez, Ann. Phys., {\bf 365}, (2016),1.\\�\\
7. G. Dattoli and A. Torre, {\it Free-electron laser theory}, CERN 89-03, March (1989).\\�\\
8. G.V. L\'opez, {\it Generalization of the Force Approach to Radiation Reaction}, arXiv:1602.03171v1, 8 Feb (2016).

\end{document}